\documentclass[sigconf,10pt]{acmart}
\usepackage{tikz}
\usepackage[english]{babel}  
\addto\extrasenglish{  
    
}
\usepackage{hyperref}  

\begin{document}

\title{Who Needs DRAM? We Have Fiber}

\author{Hannah Atmer, Yuan Yao, Thiemo Voigt, Stefanos Kaxiras}
\affiliation{%
  \institution{Uppsala University}
  \city{Uppsala}
  \country{Sweden}
}

\sloppy

\begin{abstract}
The rising pressure on DRAM availability and contract pricing reflects generative AI's massive high-performance memory requirements. This pressure is heavily compounded by hyperscale data center expansion, which now consumes a significant portion of global DRAM output. In this work, we propose a new architecture, Fiber Memory, which reimagines the role of optical fiber in a hyperscale data center, deploying it as an active, recirculating delay-line memory for immutable data, such as large language model weights. We present a hybrid data- and pipeline-parallel optical broadcast delay-line memory architecture that accounts for fiber's physical realities. By incorporating space-division multiplexed multi-core fibers, passive optical tap-and-amplify interfaces, co-packaged optics, and centralized in-ring 3R (re-amplify, reshape, retime) regeneration, our case study evaluation suggests that Fiber Memory can eliminate redundant weight storage across tens of thousands of AI accelerators and reduce weight-delivery power by over 75\% compared to traditional HBM3e configurations.
\end{abstract}






\copyrightyear{2026}
\acmYear{2026}
\setcopyright{cc}
\setcctype{by}
\acmConference[HotNets '26]{The 25th ACM Workshop on Hot Topics in Networks}{November 16--17, 2026}{Salt Lake City, UT,
USA}
\acmBooktitle{The 25th ACM Workshop on Hot Topics in Networks (HotNets '26), November 16--17, 2026, Salt Lake City, UT,
USA}
\acmDOI{10.1145/3847290.3859938}
\acmISBN{979-8-4007-3029-0/2026/11}

\maketitle

\section{Introduction and Motivation}

Memory is a bottleneck in modern computing clusters running large language models (LLMs). Traditional hardware platforms depend on stacking high-bandwidth memory (HBM) or double-data-rate (DDR) DRAM directly adjacent to processing units to feed billions of model parameters into arithmetic pipelines. This paradigm has driven a massive surge in DRAM demand, resulting in supply constraints, high costs, and thermal and power limits within hyperscale data centers~\cite{forbes}. 
In this work, 
we draw on two insights to propose a new paradigm in which the optical fiber network is used as \emph{memory} to avoid unnecessary data replication and re-fetch.

\textbf{Insight 1: Weight replication in DRAM memory across a datacenter is immensely inefficient}. Consider that in a hyperscale datacenter, the same model parameters (e.g., Attention and FFN weights) are replicated across all the nodes that serve the same LLM. Not only are model parameters replicated extensively, but \emph{far worse}, the accesses (requests-responses) to such replicated data are identically performed by every node that serves the same model. Replication leads to excessive energy consumption.

\textbf{Insight 2: Fiber is Memory}. The sheer amount of fiber in a hyperscale datacenter (10,000 to 100,000 km of fiber strand) holds an immense number of bits at any time. A simple loop of fiber that spans the datacenter has a capacity of multiple TB of data that circulate past every compute node in the loop at 2/3 the speed of light ($c$). Effectively, we can turn fiber into a \emph{Delay-line Memory}, one of the first types of memory used in electronic computers~\cite{wiki:delay_line_memory}, albeit with a tremendous speed, capacity, and length, compared to the Mercury delay-line memories of the 1940s and early 1950s.

Instead of thousands of nodes consuming local electrical energy to fetch identical weights from HBM or DDR, a single centralized optical transmitter can feed the model parameters once into the shared fiber network. Inference nodes passively ``tap'' the data circulating continuously in the fiber, extracting weights on the fly and eliminating redundant weight storage and fetch energy across the cluster. 

Our proposal is supported by the growing adoption of co-packaged optics (CPO) which places silicon photonics engines directly onto the processor substrate to bypass power-hungry electrical transceivers~\cite{fotouhi2019suma, huawei}. In CPO, the silicon switch chip and the optical engines (silicon photonics chips) are placed on the same package substrate. The electrical signal only has to travel millimeters instead of centimeters, which avoids the need for a digital signal processing unit that consumes most of the power of the optical to electrical conversion~\cite{HIRch9} and obviates the need for buffering in RAM, enabling us to feed data directly into compute units.

Existing state-of-the-art datacenter networking hardware also seeks to minimize data transfer costs but is still based on intermediary buffering. For example, NVIDIA's GPUNetIO receives the incoming network packets and streams the raw data directly into the GPU's VRAM using GPUDirect RDMA~\cite{nvidia} and some FPGA SmartNICs briefly buffer data in pipeline registers, small FIFOs, or FPGA block RAM~\cite{fpgasurvey}. In our approach, however, we strive to obviate, not only the need for storage for immutable data, but also for any buffering on their way to the computing units where they will be used.

In this work, we make the following contributions:
\begin{enumerate}
    \item Establish the feasibility of using data center fiber as a high-speed delay-line memory for AI accelerators. 
    \item Sketch an optical network architecture utilizing multi-core fibers, passive splitters, cladding-pumped multi-core optical amplifiers, and a single in-ring 3R regenerator to distribute weight streams.
    \item Detail data representation, alignment, and interleaving strategies to feed processing systolic arrays directly from fiber.
    \item Provide a quantitative evaluation of the performance, latency, and power consumption of Fiber Memory compared to standard HBM-based systems.
\end{enumerate}

\section{Fiber as Memory}

\subsection{Bit Capacity of Fiber}

The capacity of an optical fiber to store data ``in flight'' is governed by the bandwidth-delay product (BDP). The propagation speed of light in a standard silica fiber core ($v$) is given by:
\begin{equation}
v = \frac{c}{n} \approx \frac{3 \times 10^8\text{ m/s}}{1.5} = 2 \times 10^8\text{ m/s} = 200\text{ km/ms}
\end{equation}
where $n \approx 1.5$ is the refractive index of the fiber core~\cite{refractiveIndex}. The total propagation delay ($\tau$) for an optical fiber strand of length $L$ is:
\begin{equation}
\tau = \frac{L}{v}
\end{equation}
If a transmitter modulates the light at an aggregate data rate of $B$, the total volume of data $M$ stored inside the fiber strand at any single instant is:
\begin{equation}
M = B \times \tau = B \times \frac{L}{v}
\end{equation}

Aggregating the route lengths of a hyperscale data center's fiber optic cables yields an aggregate fiber strand length ($L$) ranging from $10,000$ to $100,000$ km~\cite{corning:dense}. Using ultra-dense wavelength-division-multiplexing, a single commercial fiber strand can achieve an aggregate bandwidth ($B$) of 
approximately $100\text{ Tb/s}$ ($12.5\text{ TB/s}$). Substituting these metrics into the BDP equation reveals the enormous latent storage capability of the network. For a 100,000 km aggregate fiber length, the storage capacity is:

\begin{align}
M &= 12.5\text{ TB/s} \times \frac{100,000\text{ km}}{200,000\text{ km/s}} = 6.25\text{ TB}
\end{align}
This capacity is more than sufficient to store multiple massive LLMs (such as a 1-trillion parameter model quantized to INT8 or FP16) entirely in flight.
While this example demonstrates the baseline potential of a single fiber strand, our complete Fiber Memory architecture aggregates multiple 16-core Multi-Core Fiber (MCF) cables. Each cable packs independent cores within a single physical glass cladding sharing a single protective jacket. This allows us to compress the spatial footprint of the delay line by over $90\%$~\cite{Matsui2017}, making the physical installation of e.g., a $1,000\text{ km}$ loop of 8 parallel MCF cables (as used in the case study of Section~\ref{sec:case}) feasible within datacenter setups.

\subsection{Fiber Memory Organization}
\label{sec:organization}

\begin{figure}[t]
    \centering
    \includegraphics[width=\linewidth]{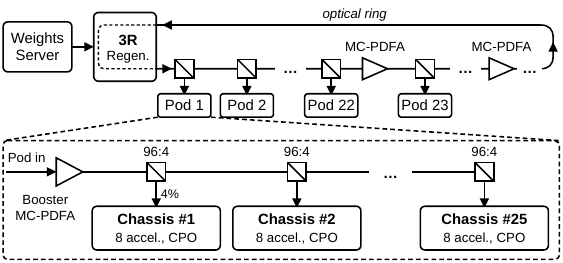}
    \caption{Ring and Pod Architecture. The weights server injects model parameters into a ring of wavelength-multiplexed, 16-core MCF cables and performs 3R regeneration once per loop circulation. Regional splitters broadcast the signal to independent pods, minimizing cumulative losses and local transceiver overhead.}
    \label{fig:bigpicture}
\end{figure}

To exploit the in-flight storage of fiber, we organize the optical network into a hierarchical ring and pod architecture (Figure~\ref{fig:bigpicture}). Traditional point-to-point network transceivers receive an optical signal, convert it to the electrical domain (O-E), process or route it, and re-modulate it back to light (E-O). This O-E-O cycle is highly energy-intensive and introduces hundreds of nanoseconds of digital processing latency. Instead, our \emph{end-to-end optical architecture} uses passive, asymmetric ``Tap-and-Amplify'' interfaces at both the ring level and the pod level.

A central weights server reads the weights from memory (e.g., multiple parallel HBM stacks) and writes them onto the fiber ring at startup. Regional passive optical splitters then broadcast the optical signals to distinct local pods of inference nodes. Inside each pod, the signal runs along an unamplified localized optical distribution bus, and passive splitters along each pod's distribution bus tap the signal directly into the pod's local nodes. We define our foundational network ``node'' not as a single monolithic chip, but as an 8-accelerator baseboard chassis (analogous to standard dense computing platforms like NVIDIA HGX). The optical signal thus reaches the co-packaged optical (CPO) receivers of the 8 accelerators of the inference chassis~\cite{fotouhi2019suma, Tan2023}. Because the distance from the chassis splitter to the silicon photodetector is mere millimeters, extraction is practically instantaneous and requires no electronic buffering.

\textbf{Signal format.} Because every chassis receives the stream
passively through a tap, the signal must be cheap to detect. We therefore
use intensity modulation with direct detection (IM-DD) instead of coherent
transmission, which avoids power-hungry coherent DSPs at every
receiver~\cite{pang}. Each MCF core carries 16 Dense Wavelength Division Multiplexing (DWDM) channels of 50\,Gbaud PAM4
(100\,Gb/s each), giving 1.6\,Tb/s (200\,GB/s) per core. The channels sit
on a 100\,GHz grid centered on the O-band zero-dispersion wavelength
($\lambda_0 \approx 1312$\,nm), which lets direct detection work over long
spans without dispersion compensation (Section~\ref{sec:challenges}).

\subsection{Signal Amplification and Regeneration}

An ideal delay-line ring would circulate light indefinitely, but standard silica fiber cores exhibit intrinsic physical attenuation that increases with the length of the fiber~\cite{Rademacher2020}. If the ring were one continuous fiber loop, then uncompensated attenuation would completely extinguish the optical signal. We therefore partition the ring into segments of length $L_S$, joined by inline optical amplifiers to restore signal amplitude. In our case study we choose a standard $50\text{ km}$ spool length in the O-band that has a manageable $\approx 16\text{ dB}$ baseline fiber loss per section~\cite{Rademacher2020}.

\begin{figure}
    \centering
    \includegraphics[width=.9\linewidth]{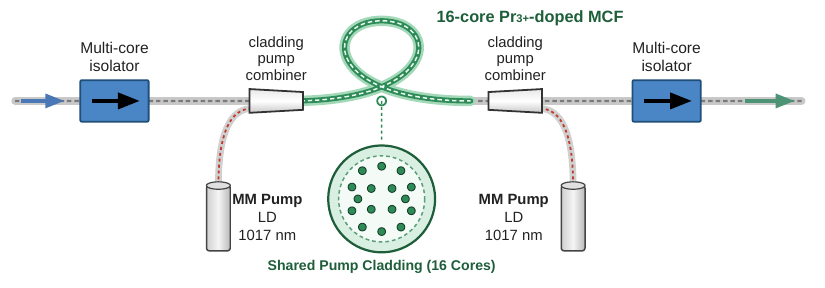}
\caption{16-Core Cladding-Pumped Praseodymium-Doped Fiber Amplifier (MC-PDFA): Bi-directional $1017\text{ nm}$ multimode pump laser diodes (MM Pump LDs) couple light into a shared microstructured cladding (inset cross-section) to excite all 16 $\text{Pr}^{3+}$-doped cores uniformly.}
    \label{fig:mc-pdfa}
\end{figure}

To amplify all cores simultaneously in the O-band, we employ a cladding-pumped Multi-Core Praseodymium-Doped Fiber Amplifier (MC-PDFA), shown in Figure~\ref{fig:mc-pdfa}, adapting the shared-cladding-pump microstructured multicore amplifier architecture demonstrated for $\text{Er}/\text{Yb}$ fibers~\cite{Zhang2024_MCFA} to a $\text{Pr}^{3+}$-doped fluoride host~\cite{Mori2021}. A shared multimode pump laser is injected into the microstructured cladding to excite all doped cores uniformly~\cite{Mori2021, Zhang2024_MCFA}. The MC-PDFA operates entirely in the optical domain with a time-of-flight latency under $75\text{ ns}$ (for a $<15\text{ m}$ doped segment)~\cite{Mori2022}, while the long upper-state lifetime of $\text{Pr}^{3+}$  ($\approx 110\,\upmu\text{s}$) renders it immune to inter-channel cross-gain modulation (XGM) and fast gain saturation~\cite{Ohishi1998}.
The MC-PDFA is conservatively assumed to provide at least $+20\text{ dB}$ gain~\cite{Nishi2022, Zhang2024_MCFA} which sets limits on how we can split (tap) the signal.

While inline MC-PDFAs offset span attenuation via linear re-amplification, analog amplification alone cannot sustain a recirculating memory indefinitely. Across endless loop circulations, ASE (amplified spontaneous emission) noise from the inline amplifiers compounds~\cite{Agrawal2012}, while residual chromatic dispersion~\cite{Wey2020} and thermal drift~\cite{Janzetal2023} introduce pulse broadening and clock timing jitter. To prevent noise and jitter from accumulating across passes, we place a single 3R regenerator~\cite{Agrawal2012} in the central ring at the weights server. At startup, the weights server injects the data onto the ring via this regenerator (Figure~\ref{fig:bigpicture}). Once every loop circulation, the 3R regenerator terminates the optical stream, re-locks the data phase via clock and data recovery, and re-modulates clean optical carriers.

\subsection{Tap and Amplify}
\label{sec:tap-and-amplify}

\noindent\textbf{Budget principle.}
Design choices in our proposal are dictated by the need to strike a balance between the number of taps and the amount of amplification. Every amplifier costs pump power, so we want as few as possible. Every tap adds loss that some amplifier down the road must recover. We take the conservative MC-PDFA gain of $+20$\,dB~\cite{Nishi2022, Zhang2024_MCFA} as our fixed budget and make one structural choice: we place the amplifiers only at the joints between fiber spools, never in the middle of a span. With standard 50\,km spools, this means exactly one 20\,dB MC-PDFA per 50\,km of ring. All remaining questions reduce to how many taps fit in what is left of that 20\,dB after we account for the attenuation. All the splitters of a particular segment are placed (in relatively close proximity) right after the front-end amplifier to tap the signal at full strength.
Each pod has its own MC-PDFA booster at its entrance, as shown in Figure~\ref{fig:bigpicture}.

\noindent\textbf{Choice of splitter.}
To tap the signal we settle on an asymmetric 4\,:\,96 passive splitter at both levels of the hierarchy. On the through path it costs only $-10\log_{10}(0.96) \approx 0.177$\,dB, so many taps can share one amplifier. The 4\% branch drops the extracted signal by $-10\log_{10}(0.04) \approx 14.0$\,dB. This is a level that a 20\,dB booster can restore (at the pod entrance) and a receiver can still detect (at a chassis). Going below 4\% can increase the number of taps per amplification, but it becomes exceedingly difficult to amplify or detect well, potentially raising the overall cost.

\noindent\textbf{Ring level: pods per segment.}
O-band single-mode cores attenuate by $\approx 0.32$\,dB/km~\cite{Rademacher2020}, so a 50\,km segment loses $16.0$\,dB to the fiber alone. That leaves $20 - 16 = 4.0$\,dB of the inline MC-PDFA's gain for regional taps. At $0.177$\,dB each, this allows at most $\lfloor 4.0/0.177 \rfloor = 22$ pod taps per segment before the next amplifier. For example, a 1,000\,km ring has twenty 50\,km segments. It needs $19 \times 8 = 152$ inline MC-PDFAs: one per segment joint for each of the 8 MCF cables. The 3R regenerator at the weights server closes the loop and provides the twentieth restoration point. It therefore supports up to $20 \times 22 = 440$ pods.

\noindent\textbf{Pod level: chassis per pod.}
Each pod has a single 20\,dB booster MC-PDFA at its entrance, directly after its regional tap. The pod bus beyond that point is fully passive. It is also quite short compared to ring segments, so we assume negligible attenuation due to fiber length.

The number of chassis we can fit in a pod without requiring re-amplification of the signal is determined by the cumulative loss on the pod bus as sequential taps are applied. 
The pod bus loss is $n \times 0.1773\text{ dB}$, where $n$ is the number of taps. The booster launches $+9$\,dBm per wavelength, so chassis $n$ receives
$9 - 14.0 - 0.177\,(n-1)$\,dBm. We can therefore attach at most 38 passive chassis before the 4\% branch hits the $-11.6$\,dBm receiver sensitivity floor. We stop at 25 chassis ($n=25$), the last one receiving $-9.25$\,dBm, a 2.35\,dB buffer for real-world fiber-to-chip coupling and micro-ring resonator (MRR) drop losses.

\noindent\textbf{Resulting scale.}
With 8 accelerators per chassis, one booster per pod and one inline amplifier per 50\,km, a 1,000\,km ring, such as the one we use for our case study in Section~\ref{sec:case}, serves up to
$20~\text{segments} \times 22~\text{pods} \times 25~\text{chassis} \times 8~\text{accelerators} = 88{,}000$ accelerators.

\section{LLM Inference at Scale}

\subsection{LLM Weight vs. Activation Footprints}

LLM execution is highly asymmetric with respect to memory access patterns. During the autoregressive decoding phase of inference, the accelerator generates tokens sequentially, one by one. For each generated token, the processor must read the complete set of model weights ($W$) from memory, while the activation data ($A$) consists predominantly of the Key-Value (KV) cache for preceding tokens.

The volume of weights vastly exceeds the size of the activations for small to medium batch sizes. In typical workloads, the model weights account for $90\%$ to $99\%$ of the total memory bandwidth consumption~\cite{Aminabadi2022}. For instance, to generate a single token in a 70-billion parameter model (FP16, 140 GB size), a GPU must load the entire 140 GB of parameters into its registers, whereas the activation and KV cache transfers total less than 1.4 GB. Because LLM inference is memory-bandwidth bound, performance is limited by how quickly the weights can be transferred from HBM. Under our paradigm, we keep the activations and KV cache in local DRAM (e.g., HBM3e) or high-density SRAM on the accelerator, while streaming $100\%$ of the massive weight footprint from the fiber.
Similarly to notable examples such as the Cerebras architecture~\cite{lie2021multiwafer}, we assume systolic-array-type accelerators that follow the activation-stationary execution model. 

\subsection{Data Alignment and Systolic Array Interfacing}

Because the weights are received continuously from the fiber, the compute units must execute matrix-vector multiplications in lockstep with the incoming light pulses. Standard processors request data using memory addresses, but Fiber Memory operates on a push-based, deterministic streaming model: the accelerator simply waits for the required layer parameters to flow through the tap.

\begin{figure}[t]
    \centering
    \includegraphics[width=0.9\linewidth]{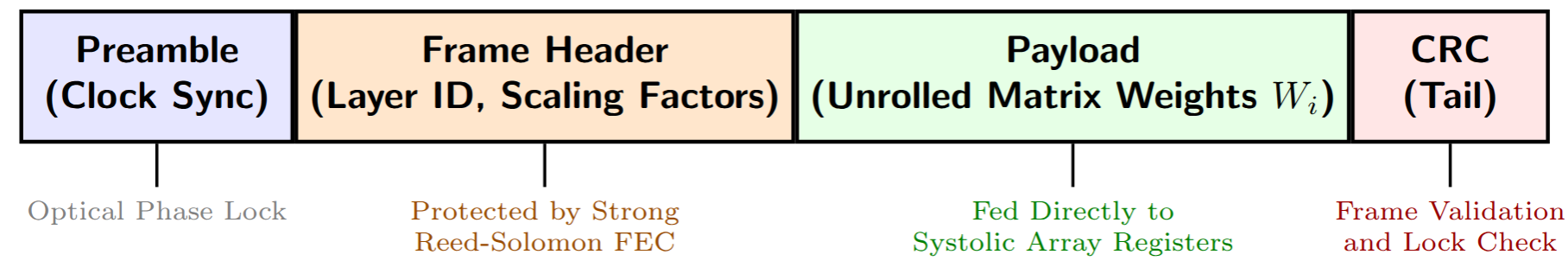}
    \caption{Structure of a streamed weight packet.}
    \label{fig:packet}
\end{figure}

To structure this stream, weights are packaged into Streamed Weight Packets (SWPs), depicted in Figure~\ref{fig:packet}.
Each SWP contains the following parts: \emph{Preamble:} A highly distinct optical pulse pattern that allows the receiver's clock and data recovery (CDR) circuits to lock onto the incoming data phase. \emph{Frame Header:} The header contains information necessary for nodes to recover from operational jitter and stalls, make use of replication, and protects the quantization scaling factors. \emph{Payload:} The raw weight matrix elements ($W_i$), formatted specifically to match the processing layout of the systolic array. \emph{CRC:} A trailing Cyclic Redundancy Check field used for final packet boundary verification.

Since the weight matrix is read directly from the tap into the execution registers without local buffering, the weights server unrolls the matrix layout beforehand, thereby eliminating the need for complex address translation, row-decoding, or layout transformations on the accelerator.

\subsection{Slack}

A major challenge in a delay-line system is coordinating the processing rate of individual nodes with the constant speed of the optical stream. If a node falls behind due to a high batch size or prolonged KV cache lookup, it might miss the beginning of the next weight packet, forcing it to wait an entire loop cycle ($\tau$) for the weights to reappear. To build robustness and scheduling flexibility, the weights server can implement padding for slack. 
As shown in Figure~\ref{fig:dp}, we can use slack spaces (high-frequency idle patterns or replicas of layers) between layer packets to give the node a safety margin for finalizing its activation and KV cache memory swaps before the next layer payload begins. 
More importantly, slack spaces need to be inserted between the Attention weights and the FFN weights of each layer (Figure~\ref{fig:dp}), since it is the Attention latency that grows with context length---fortunately, at a slow enough pace to accommodate large increases in context windows~\cite{Pope2023}.

\subsection{Execution Model}

\begin{figure}[t]
    \centering
    \includegraphics[width=\linewidth]{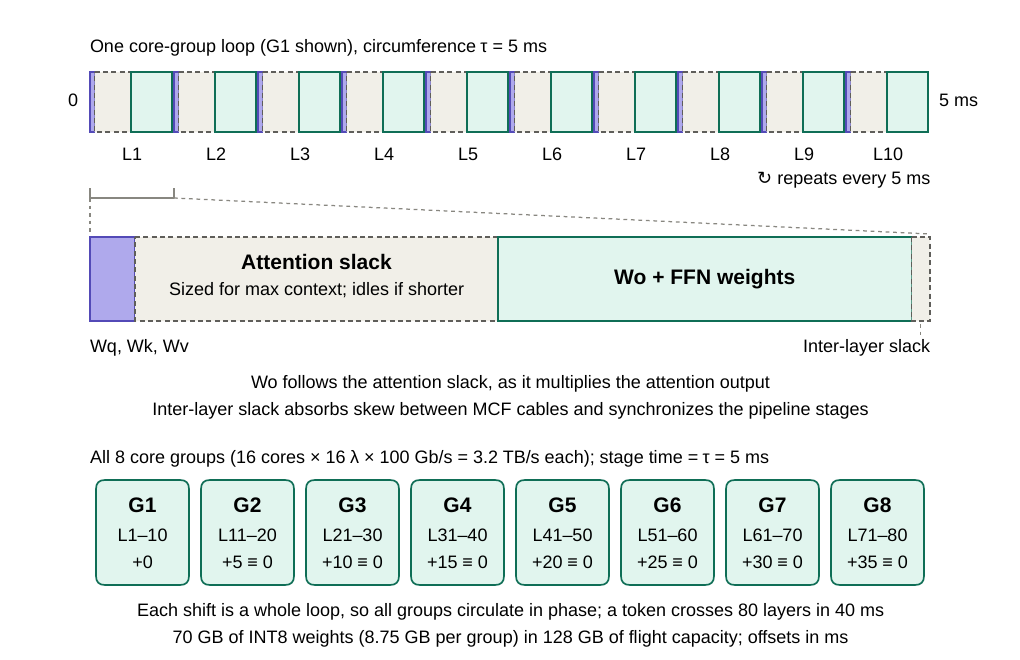}
    \caption{Layer slots in one core-group loop, and the eight time-shifted core groups. Slack inserted after Attention and between layers for pipeline synchronization.}
    \label{fig:dp}
\end{figure}

An entire chassis receives the full 8-cable MCF bundle (carrying all 128 cores), allowing the chassis to ingest the entire model stream simultaneously. This guarantees that each chassis can independently complete end-to-end inference passes without transferring activations over a backend network.

However, terminating 128 physical fibers into a single silicon chip presents severe manufacturing and thermal challenges. To address this, we physically distribute the MCF cables internally among the 8 accelerators on the chassis board. Each individual accelerator package only physically taps 16 cores, i.e., one MCF cable, in order to avoid potential thermal skew between cables. 

\begin{figure}[t]
    \centering
    \includegraphics[width=0.8\linewidth]{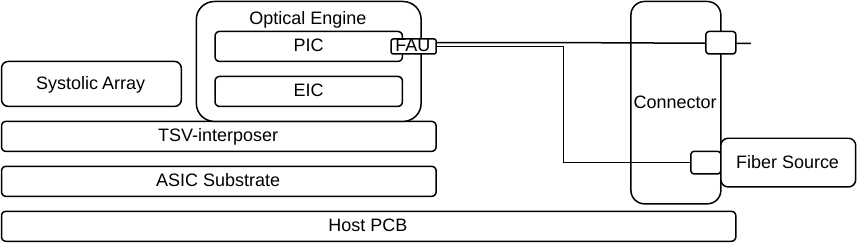}
    \caption{Co-packaged optics accelerator integration.}
    \label{fig:cpo}
\end{figure}

The Photonic Integrated Circuits (PICs) are positioned on the same interposer as the compute silicon (Figure~\ref{fig:cpo}). Inside the local PIC of each accelerator, an array of silicon MRRs demultiplexes 16 wavelengths per core and feeds raw weight parameters directly to the systolic array registers. This restricts the total on-chip receiver interface to just 256 MRRs ($16 \text{ cores} \times 16 \text{ wavelengths}$), which is highly manufacturable, thermally stable, and commercially viable using current silicon photonics packaging techniques~\cite{Tan2023}.

\begin{figure}[t]
    \centering
    \includegraphics[width=\linewidth]{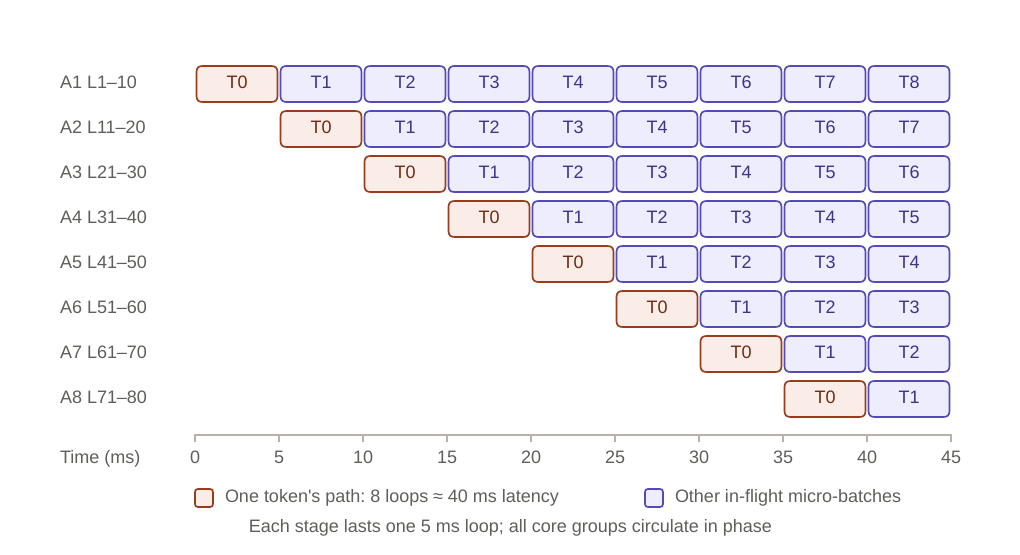}
    \caption{Pipeline occupancy across accelerators A1–A8.}
    \label{fig:pipeline}
\end{figure}

The model is split across the 8 accelerators inside a chassis. For example, for Llama-3-70B used in Section~\ref{sec:case}, A1 gets layers 1–10, A2 gets 11–20, and so on (Figure~\ref{fig:dp}). Since each accelerator ingests only 1/8th of the model, the accelerators work together as a pipeline and share a DRAM that contains the KV cache for the batch of 8 tokens currently in the pipeline (Figure~\ref{fig:pipeline}). Once an accelerator has completed processing of its contiguous block of layers, it hands off its activations over the board interconnect to the accelerator with the next contiguous block of layers for that token.

\section{Case Study: Llama-3-70B}
\label{sec:case}

To evaluate our architecture under physically consistent and realistic design constraints, we analyze a cluster deployment running a 
70-billion-parameter dense model (Llama-3-70B). 

\emph{Model Size ($W$):} $70\text{ GB}$ (INT8), with an aggregate fiber capacity scaled to $128\text{ GB}$ to accommodate timing slack spaces. The up to 45\% slack space in this design accommodates the increase in attention latency associated with longer context lengths and can be shortened for higher throughput.

\textit{Aggregate bandwidth:} Our target is the memory bandwidth of a high-end AI processing engine (e.g., NVIDIA H100 with $3.35\text{ TB/s}$ HBM3~\cite{nvidia_h100}). Recall that each MCF core carries 16 DWDM channels at $100\text{ Gb/s}$ (Section~2.2), so assigning 16 cores to each accelerator provides:
\begin{equation}
\begin{aligned}
\text{Bandwidth}_{\text{acc}}
    &= 16\,\text{cores} \times 16\,\text{wavelengths} \times 100\,\text{Gb/s} \\
    &= 25.6\,\text{Tb/s} = 3.2\,\text{TB/s}
\end{aligned}
\end{equation}
With 8 accelerators per chassis, every chassis needs 128 cores, which we provide with 8 parallel cables of 16-core MCF. The full bundle carries $204.8\text{ Tb/s}$ ($25.6\text{ TB/s}$) and is routed into every chassis, where it is partitioned among the 8 accelerators.

\textbf{Physical ring dimensioning \& delay.}
The ring must hold the $70\text{ GB}$ model plus $58\text{ GB}$ timing slack, so we target $M = 128\text{ GB}$ in flight. At the aggregate bandwidth $B = 25.6\text{ TB/s}$, the loop must delay the stream by:
\begin{equation}
\tau = \frac{M}{B} = \frac{128\text{ GB}}{25.6\text{ TB/s}} = 5\text{ ms}
\end{equation}
With light propagating at $\approx 200,000\text{ km/s}$ in silica fiber, this corresponds to a loop length of:
\begin{equation}
L = \tau \times v = 5\text{ ms} \times 200,000\text{ km/s} = 1,000\text{ km}
\end{equation}
Each of the 8 MCF cables is therefore built from 20 cascaded stages of standard commercial $50\text{ km}$ spools.

\emph{Cluster Size:} 
Out of a maximum possible 88,000 accelerators that can fit in this topology (Section~\ref{sec:tap-and-amplify}), we conservatively model 10,000 accelerators (1,250 8-accelerator chassis). The 1,250 chassis are organized into 50 Pods of 25 chassis each (200 accelerators per Pod). The 50 Pods are spread across the 20 segments.

\noindent\textbf{Baseline Energy and Power Projections.}
We compare our direct-detection model against a standard HBM3e baseline. We deliver a weight throughput of $25.6\text{ Tb/s}$ per accelerator comparable to an AI accelerator package equipped with two $1.6\text{ TB/s}$ HBM3e~\cite{micron:hbm3e}. Using an energy metric of $4.0\text{ pJ/bit}$ for HBM3e local memory fetches~\cite{Lee2024}, the weight-delivery power for 10,000 traditional HBM3e cluster accelerators is:
\begin{align}
P_{\text{HBM}} &= 10,000 \times 25.6\text{ Tb/s} \times 4.0\text{ pJ/bit} = 1024\text{ kW}
\end{align}

This 1024 kW calculation represents a theoretical maximum assuming continuous, peak-bandwidth utilization. Although practical inference workloads experience computational micro-stalls that slightly reduce dynamic fetch rates, we deliberately omit the substantial static leakage and refresh power overhead inherent to massive HBM3e clusters, keeping this baseline conservatively balanced. Furthermore, while Fiber Memory eliminates the HBM fetch energy for weights, nodes will still expend a small fraction of local energy accessing on-chip SRAM or high-density DRAM for the remaining 1\% to 10\% of memory bandwidth required by activations and the KV cache.

\noindent\textbf{Fiber Memory Energy and Power Projections.}

\emph{Central Transmitter \& In-Ring 3R Regenerator:} 
We modulate 128 active cores $\times$ 16 channels = 2,048 total lasers. At $100\text{ mW}$ optical power per laser and an O-band Distributed Feedback (DFB) laser wall-plug efficiency of $5\%$~\cite{Guler2023}, the central laser source draws $2,048 \times 2\text{ W} \approx 4.1\text{ kW}$. At the end of the $1,000\text{ km}$ loop, a single central 3R regenerator terminates the $204.8\text{ Tb/s}$ stream via a 2,048-channel direct-detection receiver and CDR retiming front-end ($204.8\text{ Tb/s} \times 0.7\text{ pJ/bit} \approx 0.14\text{ kW}$~\cite{Ahmed2021}) before re-modulating the central lasers:
\begin{equation}
P_{\text{central\_3R}} = 4.1\text{ kW} + 0.14\text{ kW} \approx 4.2\text{ kW}
\end{equation}

\emph{Central Delay Line Amplification:} The $1,000\text{ km}$ loop requires inline amplification between each $50\text{ km}$ spool to prevent total signal extinction. With 8 MCF cables and 20 stages, we deploy 160 (152 plus 8 at the 3R site) inline 16-core cladding-pumped MC-PDFAs. Based on the $\approx 26\text{ W}$ multimode optical pump power required to simultaneously excite all cores in a microstructured multicore amplifier~\cite{Zhang2024_MCFA} and a $50\%$ pump diode wall-plug efficiency, each 16-core MC-PDFA draws $50\text{ W}$ of power:
\begin{equation}
P_{\text{loop\_amps}} = 160 \times 50\text{ W} = 8.0\text{ kW}
\end{equation}

\emph{Pod Booster MC-PDFAs:} With 50 Pods routing the full 8-cable bundle and using $4:96$ chassis taps, we require only 1 booster MC-PDFA per cable at the entrance of each Pod and no inline Pod amplification:
\begin{equation}
P_{\text{pod\_amps}} = 50\text{ Pods} \times 8\text{ cables} \times 50\text{ W} = 20.0\text{ kW}
\end{equation}
        
\emph{Local IM-DD PIC Receivers, Equalizers \& FEC:} By utilizing direct-detection PAM4 instead of coherent technology, we bypass power-hungry ADCs and massive coherent DSPs. The short-reach receiver electronics operate at an aggregate $0.7\text{ pJ/bit}$~\cite{Ahmed2021}:
\begin{align}
P_{\text{receivers}} &= 10,000 \times 25.6\text{ Tb/s} \times 0.7\text{ pJ/bit} = 179.2\text{ kW}
\end{align}
\nopagebreak
Summing these terms yields the total Fiber Memory power consumption:
\begin{equation}
P_{\text{fiber\_total}} = 4.2\text{ kW} + 8.0\text{ kW} + 20.0\text{ kW} + 179.2\text{ kW} = 211.4\text{ kW}
\end{equation}

Comparing the baseline and fiber models, Fiber Memory achieves a 79.4\% reduction in total weight-delivery power ($211.4\text{ kW}$ vs. $1,024\text{ kW}$) across the cluster, while completely eliminating the static leakage power, cooling overhead, and high capital expense of storing $87.5\text{ TB}$ of replicated static model weights across 1,250 pipelined chassis.

\emph{Energy per token:} Since the optical infrastructure operates continuously, energy per token scales inversely with useful throughput. Accelerators operate at bandwidth saturation and are power-gated when idle, so to first order both unused slack and missed deadlines reduce useful throughput. The two are opposing forces. Slack exists to prevent missed deadlines, so minimizing slack reduces idle time but increases missed deadlines. Deploying separate rings with different slack budgets for different classes of context lengths addresses both, since each request is dynamically routed to a ring whose slack matches its needs. This keeps both idle time and missed deadlines low, and thus minimizes energy per token. Future work will delve deeper into these tradeoffs.

\section{Challenges and Physical Constraints}
\label{sec:challenges}

Deploying a multi-kilometer recirculating fiber memory requires careful management of the physical limitations of fiber optics.

\noindent\textbf{Chromatic Dispersion Management.}
To avoid excessive chromatic dispersion, our architecture localizes operations within the O-band's zero-dispersion regime. We deploy a 16-channel DWDM grid with $100\text{ GHz}$ spacing, confining the entire transmission within a narrow $1500\text{ GHz}$ ($\sim 8.6\text{ nm}$) spectral window centered exactly at $\lambda_0 = 1312\text{ nm}$. By restricting the total spectral footprint and centering it precisely at the fiber's zero-dispersion wavelength, we bound the physical dispersion coefficient at $|D| \leq 0.4\text{ ps/(nm}\cdot\text{km)}$ across all active channels, minimizing pulse broadening without the need for active dispersion compensation~\cite{Wey2020}.

\noindent\textbf{Forward Error Correction and LLM Error Tolerance.}
To maximize the distance between optical regenerators, we utilize high-throughput Forward Error Correction integrated directly into the accelerator's PIC receiver. Low-overhead Reed-Solomon or Low-Density Parity-Check codes can correct raw input Bit Error Rates as high as $10^{-3}$ down to a clean $10^{-12}$ operating standard, with minimal latency overhead ($<100\text{ ns}$~\cite{Chang2022}).
LLMs also tolerate minor errors. Quantized weights are robust to random bit flips provided scaling factors and layer normalizations remain intact~\cite{Gao}. By strongly protecting SWP headers and scaling factors, minor residual payload errors can be tolerated, relaxing optical signal-to-noise ratio (OSNR) and transceiver requirements.

\noindent\textbf{Thermal Management.}
Silicon photonics chips are highly sensitive to thermal variations. MRRs rely on sub-micron dimensions to achieve wavelength resonance; a minor temperature shift drifts the refractive index, causing the MRR to miss its target wavelength channel~\cite{Janzetal2023}. Active thermal tuning is relegated to micro-thermoelectric heaters integrated into the MRRs, which consume only micro-watts per channel to stabilize the optical alignment~\cite{Tan2023}. 
Laser diodes are the most vulnerable component in any optical system. High-power laser diodes operated on hot processor interposers degrade rapidly, presenting a severe reliability risk~\cite{Buscaino2021}. Fortunately, this is not an issue in our design as all lasers can be housed far from the hot accelerator chips. 

\section{Severing the Ring}
Looking forward, the interesting question is what would change if we sever the ring and end up with an open-ended fiber-optic broadcast line instead. Both topologies share the technical stack that makes them viable: weight broadcast, optical tapping and amplification, co-packaged optics delivering weights directly to accelerator registers without intermediate buffering (other than buffering needed to address skew). 
In a ring topology, the full model is read once and then re-circulated multiple times without active use of DRAM/HBM. In contrast, in an open-ended line architecture, the full model needs to be read from HBM for every broadcast. However, this also means that the slack can be easily re-sized for each broadcast, by appropriately timing the reads.
Lifting the ring length restrictions also helps better manage thermal and length skew among the MCF cables (as they can be much shorter than before) and expands the choices for the execution model that can be used. However, the architectural restriction on accelerators (activation-stationary since the weights are streaming) remains for both the ring and the open-ended line topologies.

\section{Conclusion}
We present Fiber Memory, a novel architecture that repurposes datacenter fiber networks as active, looping delay-line memory systems. By streaming static, highly replicated LLM weights continuously over parallel space-division multiplexed multi-core fibers we bypass the need for massive, redundant localized HBM or DDR memory blocks at every compute node. 
Our analysis indicates that our design is a reasonable starting point for developing an experimental prototype. By delivering identical weight parameters to thousands of inference pipelines simultaneously, Fiber Memory could cut weight-delivery power by over 75\% compared to HBM3e while reducing data-replication costs. Scaling Fiber Memory to larger models requires increasing the number of 50 km fiber segments and inline MC-PDFAs to accommodate a larger amount of data circulating in the loop. 

\begin{acks}
This work was supported by the
\grantsponsor{bos}{Swedish Foundation for Strategic Research}{https://strategiska.se/en/} 
grant
\grantnum{bos}{FUS21-0067} 
and by the 
\grantsponsor{vrstg}{Swedish Research Council (VR)}{https://www.vr.se/} under Starting Grant 
\grantnum{vrstg}{2025-04436}.
Generative AI tools (Gemini, Claude) were used to refine text and figures.
\end{acks}

\bibliographystyle{ACM-Reference-Format}
\bibliography{references}

\end{document}